\documentclass[useAMS,usenatbib]{mn2e}

\usepackage{graphicx}

\title[Double-mode radial--non-radial RR~Lyrae stars]{Double-mode radial--non-radial RR~Lyrae stars in the OGLE photometry of the Galactic bulge}
\author[H. Netzel, R. Smolec \& P. Moskalik]
{H. Netzel$^{1}$\thanks{E-mail: henia@netzel.pl},
R. Smolec$^{2}$\thanks{E-mail: smolec@camk.edu.pl} and
P. Moskalik$^{2}$\\
$^{1}$Instytut Astronomiczny, Uniwersytet Wroc\l{}awski, ul. Kopernika 11, 51-622 Wroc\l{}aw, Poland\\
$^{2}$Nicolaus Copernicus Astronomical Centre, Polish Academy of Sciences, Bartycka 18, 00-716 Warszawa, Poland\\
}

\begin{document}

\date{Accepted . Received ; in original form }

\pagerange{\pageref{firstpage}--\pageref{lastpage}} \pubyear{2014}

\maketitle

\label{firstpage}

\begin{abstract}
Non-radial modes are excited in classical pulsators, both in Cepheids and in RR~Lyrae stars. Firm evidence come from the first overtone pulsators, in which additional shorter period mode is detected with characteristic period ratio falling in between $0.60$ and $0.65$. In the case of first overtone Cepheids three separate sequences populated by nearly $200$ stars are formed in the Petersen diagram, i.e. the diagram of period ratio versus longer period. In the case of first overtone RR~Lyrae stars (RRc stars) situation is less clear. A dozen or so such stars are known which form a clump in the Petersen diagram without any obvious structure. 

Interestingly, all first overtone RR~Lyrae stars for which precise space-borne photometry is available show the additional mode, which suggests that its excitation is common. Motivated by these results we searched for non-radial modes in the \mbox{OGLE-III} photometry of RRc stars from the Galactic bulge. We report the discovery of 147 stars, members of a new group of double-mode, radial--non-radial mode pulsators. They form a clear and tight sequence in the Petersen diagram, with period ratios clustering around $0.613$ with a signature of possible second sequence with higher period ratio ($0.631$). The scatter in period ratios of the already known stars is explained as due to population effects. Judging from the results of space observations this still mysterious form of pulsation must be common among RRc stars and with our analysis of the OGLE data we just touch the tip of the iceberg.
\end{abstract}


\section{Introduction}
 RR~Lyrae stars are large amplitude pulsators, oscillating with periods between $\sim\!0.3$ and $1$\thinspace day. Majority of these stars are radial mode pulsators pulsating either in the fundamental mode (F mode, RRab stars) or in the first overtone (1O mode, RRc stars). Altogether, only the OGLE Catalog of Variable Stars \citep[OIII-CVS,][]{ogleIII,ogle_rr_lmc,ogle_rr_smc,ogle_rr_blg} contains nearly $43\,000$ single-mode RR~Lyrae pulsators. Double-mode pulsators, pulsating simultaneously in the F and 1O modes (RRd stars) are less frequent (1335 in OIII-CVS). Very scarce are double-mode pulsators pulsating simultaneously in the F and 2O modes. Only 17 such object are known and they were discovered only very recently \citep[see a review by][]{pam13,benko14}.

RR~Lyrae stars are very important in the broad astrophysical context. They are excellent distance indicators and serve as tracers for studies of the Galactic kinematics, structure and evolution \citep[e.g.][]{smith}. Despite their importance, the pulsation properties of RR~Lyrae stars are not well understood. The most puzzling is the Blazhko effect -- a long-term quasi-periodic modulation of pulsation amplitude and phase \citep[see e.g.][]{szabo14}. Analysis of space-borne photometry \citep{benko10,benko14} and top quality ground-based photometry \citep{jurcsik} indicates that nearly half of the RRab stars display the effect. Frequency of the effect is likely lower in the case of RRc stars. The ground-based data indicate that up to $10$ per cent of these stars show the effect \citep[e.g.][]{mizerski, nagy}. Unfortunately no Blazhko RRc star was observed from space yet. The excellent {\it Kepler} and {\it CoRoT} observations led to progress in our understanding of the Blazhko effect, however its origin remains a mystery \citep[for a review see][]{szabo14}.

The other puzzle is the excitation of non-radial modes in the first overtone RR~Lyrae stars. In their analysis of {\it MOST} photometry for RRd star AQ~Leo, \cite{aqleo} found additional mode with period shorter than the first overtone period, ratio of the two periods is $P_{\rm X}/P_{1}=0.6211$. \cite{om09} detected additional mode in 6 RRc stars of $\omega$~Centauri with similar strange period ratios (all in a range $0.608-0.622$). Period ratios in this range are far from the expected period ratios for radial modes. Four additional RRc stars with the same mysterious period ratio  were found in the OGLE LMC data \citep{ogle_rr_lmc} and in the SDSS data \citep{sdss}. In all these stars the additional mode has a very small amplitude, in the mmag regime. Interestingly, in majority of the stars observed from space period doubling of additional mode was detected (\cite{pamsm14} and references therein, \cite{szabo_corot}). Analysis of the superior quality {\it Kepler} and {\it CoRoT} photometry for RRc stars points that excitation of this mysterious additional mode in RRc stars might be common. All four RRc stars observed with {\it Kepler} show additional mode with the discussed period ratio \citep{pam_rrc,pamsm14}. In addition three stars observed with {\it CoRoT} (2 RRc and one RRd) also show the additional periodicity \citep{szabo_corot,chadid}. Altogether 18 RRc stars are known with the additional mode excited. They form a clear group in the Petersen diagram and as such are now regarded as a new group of radial--non-radial double-mode pulsators \citep[see][]{pamsm14,pam14}.

Interestingly, a very similar phenomenon is observed in first overtone Cepheids \citep[for a summary see][]{pam14}. In nearly $200$ of these stars additional periodicities, with period ratios in a range $0.60-0.65$, were found. They form three well separated sequences in the Petersen diagram \citep[see also fig. 2 in][]{pam14}. For both RR~Lyrae stars and Cepheids we do not understand the nature of the additional mode. It must be a non-radial mode \citep[][see also discussion in Sec.~\ref{sec.discussion}]{pamsm14}, however, its identification and excitation mechanism behind are not known \citep[see][]{wd12,wdrs09}. We note that in the first overtone Cepheids yet another group of non-radial modes was detected, periods of which are close to the first overtone period \citep{mk09}. 

Motivated by the {\it Kepler} and {\it CoRoT} results for RRc stars, suggesting that excitation of the additional mode might be a common phenomenon, we conducted a search for additional periodicities in the ground-based photometry of RRc stars. The publicly available photometry of the OIII-CVS is the best source for this search, as it offers a long time coverage, high photometric accuracy and large number of RRc stars. In addition we decided to focus our search on the Galactic bulge RRc stars only \citep{ogle_rr_blg}, which are much brighter than Magellanic Clouds' stars. Consequently, the noise level is lower and chances for detecting additional low-amplitude periodicities are largest. 

The data were downloaded from OGLE-III on-line Catalog of Variable Stars. The catalog contains photometry, obtained with the difference image analysis (DIA), collected during 1997--2009 years. We selected photometry in the $I$-band for all RRc and RRd stars from the Galactic bulge (5080 stars). The data cover magnitude range 13.055--19.649 mag. Number of data points per star vary from 589 to 4425 with modal value of 2830.

Our search resulted in discovery of $145$ new RRc stars and two RRd stars with additional mode excited, of which 83 stars are firm detection and the remaining stars are strong candidates. Thus the number of these interesting stars is significantly increased  allowing a more detailed study of the phenomenon. 

In Section~\ref{sec.methods} we describe the semi-automatic procedures for searching the additional periodicities. Results are presented in Section~\ref{sec.results} and discussed in detail in Section~\ref{sec.discussion}. Our most important findings are summarised in Section~\ref{sec.summary}. In the Appendix we provide a list and detailed notes on other interesting stars with additional periodicities we have found during this study.

\section{Search for additional mode in RRc stars}\label{sec.methods}

The OIII-CVS contains 4989 RRc stars in the Galactic bulge. That is why we had to develop a good automatic method for searching for stars with interesting non-radial mode. In a nutshell, the adopted procedure combines Fourier analysis of the original data, which allows to find period of the first overtone ($P_{\rm 1}$), prewhitening the data with the Fourier series and Fourier analysis of the residuals. At this stage analysis was performed over a frequency range from 0 to 10 ${\rm d^{\rm -1}}$. Based on the Petersen diagram for the already known first overtone pulsators with non-radial mode \citep[fig. 2 in][]{pam14} we assume that the period of the additional mode, $P_{\rm X}$, should fall in a range  $\left \langle 0.58, 0.64 \right \rangle P_{\rm 1}$. Corresponding frequency range was searched. In order to decide whether a signal is significant we used arbitrarily chosen criterion of signal-to-noise ratio, $S/N \geqslant 4$. 

Besides the non-radial mode in question, there are many more possible signals in the data which hamper the automatic analysis. Additional signals usually come from slow trend in the data, long-term modulation of pulsation amplitude and phase (the Blazhko effect), period change, instrumental effects or other radial modes. Slow trend produces peaks in the low frequency domain of the spectrum. The Blazhko effect typically manifests as equally spaced multiplets at the main frequency and its harmonics. Period change or long-period Blazhko effect lead to unresolved signal at $f_{\rm 1}$ and its harmonics. Periods of other radial modes form characteristic period ratios with the first overtone period different than $\left \langle 0.58, 0.64 \right \rangle$. Quite often we detect additional peaks at close-to-integer frequency values, typically the highest peak at ${\rm 2 d}^{-1}$ and its aliases. We interpret these signals as of instrumental origin.

Peaks associated with most of the discussed effects should not fall in the frequency range of interest, however their daily aliases -- inherent to ground-based photometry -- can. Therefore, we need to consider every signal before searching for the additional mode.

Trends were modelled with additional sine function of long period ($50\,000$ days) fitted to the data. The procedure works well for linear and slow parabolic trends but in many cases signal in low frequencies remained and produced aliases. During further analysis remaining signal, its daily aliases, and neighbouring $\pm 2/T$ frequency range were excluded from the automatic search. Here $1/T$, where $T$ is the data length, is a formal resolution of the Fourier transform. In the following analysis however, we adopted a conservative criterion and regard two frequencies as resolved if their separation $\Delta f > 2/T$.

We cannot model the instrumental signal ($\sim 2\thinspace{\rm d}^{-1}$) so we only excluded it and its aliases from further analysis.

For many stars, after prewhitening the data with primary frequency and its harmonics, we were able to find signal in the vicinity of $f_{\rm 1}$. In the case of unresolved signal, which may correspond to period change or long-period Blazhko effect, we excluded its daily aliases from analysis. Resolved frequencies were assumed to originate from the Blazhko effect with modulation period $1/\Delta f$. We note that search for additional signal at $f_{\rm 1}$ was restricted to $\left \langle f-0.2,f+0.2 \right \rangle$ range which was dictated by the shortest Blazhko period known \citep[above 5 days,][]{skarka}. We prewhitened the data with multiplet components, $kf_{\rm 1} \pm \Delta f$, and the whole procedure was repeated from the beginning, i.e. new residual data were inspected for the presence of a signal at low frequencies and of instrumental effects, and finally again for a signal near the primary frequency. Described method should not be considered as an analysis of the Blazhko effect, but rather as a way to clear the spectrum from unwanted signals, without proper distinction for origin of the signal. For the adopted procedure there was no difference between the real Blazhko modulation and alias of other signal accidentally placed close to the main frequency. For this reason we are not able to say precisely how many RRc stars show the Blazhko effect. This issue will be considered in a forthcoming publication. When no significant peak in the proximity of $f_1$ was found, residuals were ready for search of the non-radial mode.

Described procedure was tested on the sample of more than 200 stars, manually analysed before. It found all real candidates from the test sample, although it sometimes misclassified stars without the non-radial mode as candidates. In these cases, the false detection was caused e.g. by the non-stationary signals present in the data (and their aliases) or by signals of unknown origin (see Appendix). Finally, the described method was applied to all RRc stars from the Galactic bulge.

All stars selected as candidates by automatic procedure were analysed manually. Frequencies found during this approach were fitted to the original data in the form of 
\begin{equation}
 m(t)=A_0 + \sum_{k=1}^{N} A_k \sin(2\pi k f t + \phi_k ),
\end{equation}
where $N$ is the order of decomposition chosen so that $A_{k}/\sigma_{k} > 4$ for each $k$. Then we performed data clipping, so all deviating points were removed. The chosen criterion was $4 \sigma$, where $\sigma$ is the dispersion of the fit.

Last step of the candidate selection procedure was to check all stars for a possible contamination. In order to do this, for each candidate we found all variable stars from OIII-CVS within 1 arcsec radius around the candidate, and checked whether their periods coincide with the secondary periodicity of the candidate star or with its aliases. As a result, we excluded one star from our sample, OGLE-BLG-RRLYR-12902, in which additional periodicity may be caused by nearby RRc star OGLE-BLG-RRLYR-12836. We note that for other candidates we cannot exclude the possibility of contamination entirely, because additional variable star may be too close to the candidate to be resolved or it is not included in the OGLE catalog.

\begin{figure}
\centering
\resizebox{\hsize}{!}{\includegraphics{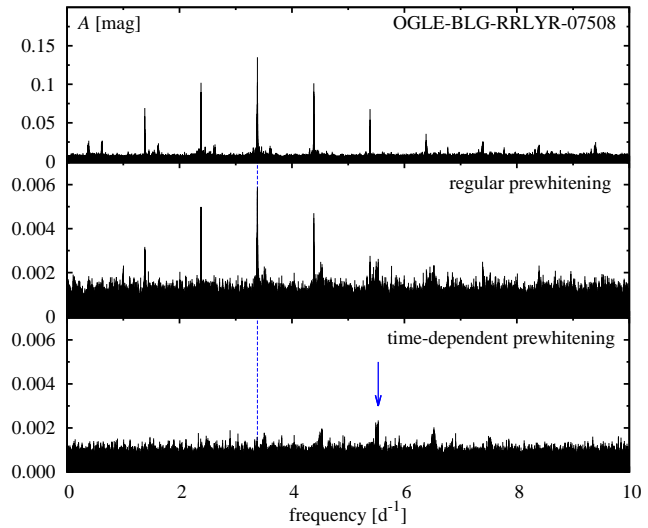}}
\caption{Illustration of the time-dependent prewhitening method. In the upper panel we plot the Fourier transform of the original data. Middle panel presents the Fourier transform after prewhitening with the primary frequency and its harmonics; unresolved signal at $f_{\rm 1}$ remains producing strong daily aliases. Lower panel shows the effects of application of the time-dependent prewhitening. Unresolved signal is removed and additional signal appears.}
\label{fig.tdfd}
\end{figure}

Strong period change is quite common among RRc stars (see next Section, and Tabs.~\ref{table:pewne} and \ref{table:kandydatki}). Consequently, in many stars we found unresolved signal at $f_{\rm 1}$. This signal and its aliases not only complicate the automatic analysis as described above, but also significantly increase the noise level in the Fourier transform and may hide the additional peak. In an attempt to remove this troublesome signal we applied a variant of a method called time-dependent prewhitening described by \cite{pamsm14}. For each star we divided the data into subsets. The division is naturally provided by observational seasons. For each subset we fitted the Fourier series with fixed frequency, derived from the whole dataset, adjusting amplitudes and phases only. Consequently, possible period changes were reflected in the seasonal variations of the Fourier phase $\phi_{\rm 1}$. Residuals from all the subsets were joined and subjected to Fourier analysis. The method works well if period change is slow, i.e. period may be regarded approximately constant within each season. In majority of the cases we could entirely remove the residual signal at $f_{\rm 1}$. We note that the method also removes seasonal zero-point differences, possibly present in the data, and non-stationarities connected e.g. to amplitude changes. As a result, noise level in the Fourier transform was decreased and we could detect additional non-radial mode in 17 stars. In Fig.~\ref{fig.tdfd} we illustrate the effects of applying time-dependent prewhitening to star with strong period change. Clearly, the noise level was significantly decreased and additional non-radial mode showed up.

In addition, we manually analysed 91 RRd stars from the Galactic bulge. We searched for the additional mode in the frequency spectrum prewhitened with frequency of the fundamental mode, first overtone and their detectable combinations.

Analysis described in this paper was conducted with dedicated software written by the authors.

\section{Results}\label{sec.results}

\begin{figure}
\centering
\resizebox{\hsize}{!}{\includegraphics{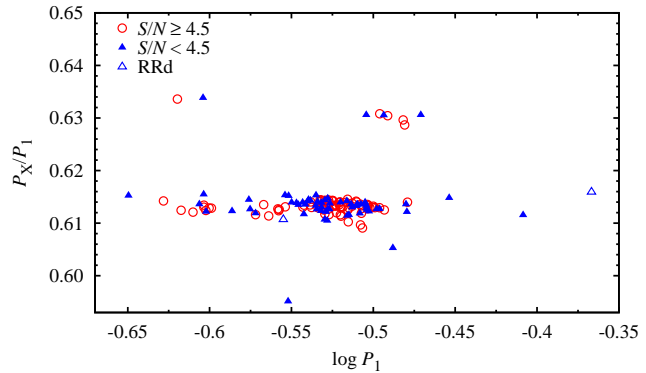}}
\caption{The Petersen diagram for newly discovered stars with additional periodicity.}
\label{fig.new}
\end{figure}

We have discovered 147 stars with additional periodicity in the frequency range on interest. In Tab. \ref{table:pewne} we listed stars with signal-to-noise ratio greater than 4.5, so we consider those stars as firm detections. There are 83 such stars. Tab. \ref{table:kandydatki} contains 64 stars with lower signal-to-noise ratio, $4 < S/N < 4.5$, so they are possible candidates for stars with additional mode. Subsequent columns contain period of the first overtone and of the additional mode, their ratio, amplitude of $f_{\rm 1}$, $A_{\rm 1}$, and $A_{\rm x}/A_{\rm 1}$. Two last columns contain signal-to-noise ratio and remarks. The Petersen diagram with all new stars is presented in Fig.~\ref{fig.new}.

The automatic procedure described in Sec. \ref{sec.methods} found 128 stars. Additional 17 stars were found after the time-dependent prewhitening was applied. They are marked with `d' in the `remarks' column of Tabs.~\ref{table:pewne} and \ref{table:kandydatki}. In addition, 67 stars for which we detected unresolved signal at $f_{\rm 1}$, a signature of a possible period change, are marked with `a'. For four stars we see a signature of the Blazhko effect and we marked them with `b'. 

For two stars with two close frequencies (dublet), which are both visible after prewhitening with $f_{\rm 1}$ and its harmonics, we included two possible solutions in the tables (in the Petersen diagrams we always plot the additional mode of the highest amplitude). In 34 stars, marked with `c', either $f_{\rm X}$ is non-stationary or additional close, but resolved signals appear after prewhitening with $f_{\rm X}$, which may also be non-stationary (see Sec.~\ref{sec.discussion}). 

In four stars, marked with `s', we see additional signal close to $1/2f_{\rm X}$, which we identify as sub-harmonic of $f_{\rm X}$. In five stars, marked with `e', we detect additional significant signal that cannot be identified with combination frequency or with sub-harmonic of $f_{\rm X}$. These stars are discussed in more detail in the Appendix.

Only in 5 stars, marked with `g' in Tabs.~\ref{table:pewne} and \ref{table:kandydatki}, we possibly see combination frequencies of the additional mode and the first overtone ($f_{\rm 1} + f_{\rm X}$), but they are very weak ($3<S/N<4$). Presence of combination frequencies proves that the two periodicities originate from the same star. In all other stars we do not see the combination frequencies. In principle they could all be blends. This is unlikely however. First, we checked for contaminations within OIII-CVS and found only one (see Sec.~\ref{sec.methods}). Second, it is statistically not possible that in all cases the period of the contaminating source has nearly the same period ratio with $P_{\rm 1}$, as is clearly visible in Fig. \ref{fig.new}. The presence of a well defined and tight sequence in the Petersen diagram is a strong argument that in all our stars we see a signature of the same phenomenon, a double-mode pulsation with first overtone and unidentified mode simultaneously excited. The same argument shows that majority of our candidate stars from Table \ref{table:kandydatki} exhibit the same form of double-mode pulsation. They perfectly fit to the sequence defined by stars from Table \ref{table:pewne} as is well visible in Fig.~\ref{fig.new}. Using the same reasoning in several stars we identified a lower amplitude alias in the frequency spectrum as a real signal, even if its daily alias was slightly higher. There were 14 such stars and they are marked with `f' in the `remarks' column of Tabs.~\ref{table:pewne} and \ref{table:kandydatki}.

Among 91 RRd stars we detected two candidates for stars with non-radial mode. They are included in Tab. \ref{table:kandydatki} with `RRd' in `remarks' and plotted in Fig. \ref{fig.new} with different symbol. They fit the main sequence very well.

\begin{table*} 
 \centering
 \begin{minipage}{140mm}
  \caption{Properties of stars with non-radial mode, detected with $S/N \geq 4.5$. Consecutive columns provide: periods of the first overtone and of additional non-radial mode, period ratio, $I$-band amplitude of the first overtone and amplitude ratio, signal-to-noise for non-radial mode and remarks.}
  \label{table:pewne}
  \begin{tabular}{@{}lccccccc@{}}
  \hline
Name & $P_{\rm 1}$\thinspace[d] & $P_{\rm X}$\thinspace[d] & $P_{\rm X}/P_{\rm 1}$ & $A_{\rm 1}$\thinspace[mag] & $A_{\rm X}/A_{\rm 1}$ &  $S/N$ & Remarks            \\
 \hline
OGLE-BLG-RRLYR-02251 & 0.24005 & 0.15210 & 0.63361 & 0.05348 & 0.066 & 4.79 & c  \\
OGLE-BLG-RRLYR-04031 & 0.31145 & 0.18970 & 0.60908 & 0.12635 & 0.026 & 4.51 & a  \\
OGLE-BLG-RRLYR-04149 & 0.25183 & 0.15433 & 0.61285 & 0.10065 & 0.043 & 4.77 & a  \\
OGLE-BLG-RRLYR-05301 & 0.30561 & 0.18730 & 0.61287 & 0.12429 & 0.039 & 4.81 & b,c  \\
OGLE-BLG-RRLYR-05311 & 0.29733 & 0.18251 & 0.61382 & 0.11748 & 0.050 & 4.67 &   \\
OGLE-BLG-RRLYR-05816 & 0.30429 & 0.18682 & 0.61395 & 0.12111 & 0.028 & 4.8 & a  \\
OGLE-BLG-RRLYR-05837 & 0.30154 & 0.18515 & 0.61401 & 0.11892 & 0.033 & 6.87 & c  \\
OGLE-BLG-RRLYR-05931 & 0.30141 & 0.18491 & 0.61348 & 0.12798 & 0.034 & 4.91 & f  \\
OGLE-BLG-RRLYR-05964 & 0.29994 & 0.18394 & 0.61324 & 0.13190 & 0.024 & 4.86 & a,d  \\
OGLE-BLG-RRLYR-06085 & 0.30688 & 0.18819 & 0.61322 & 0.13078 & 0.036 & 5.19 & g  \\
 & 0.30688 & 0.19084 & 0.62188 & 0.13075 & 0.030 & 4.16 & f  \\
OGLE-BLG-RRLYR-06352 & 0.31627 & 0.19394 & 0.61321 & 0.11370 & 0.023 & 4.69 & a,d,c  \\
OGLE-BLG-RRLYR-06374 & 0.27296 & 0.16688 & 0.61137 & 0.10634 & 0.033 & 4.58 &   \\
OGLE-BLG-RRLYR-06383 & 0.31355 & 0.19238 & 0.61355 & 0.12380 & 0.025 & 5.19 & a,d,c  \\
OGLE-BLG-RRLYR-06420 & 0.31401 & 0.19262 & 0.61342 & 0.11398 & 0.051 & 4.94 & c,f  \\
OGLE-BLG-RRLYR-06439 & 0.30499 & 0.18743 & 0.61454 & 0.12127 & 0.039 & 5.85 & a,c  \\
OGLE-BLG-RRLYR-07094 & 0.26797 & 0.16389 & 0.61161 & 0.11120 & 0.036 & 4.7 & a,d  \\
OGLE-BLG-RRLYR-07118 & 0.31578 & 0.19357 & 0.61298 & 0.11950 & 0.024 & 5.01 & a,d  \\
OGLE-BLG-RRLYR-07256 & 0.25101 & 0.15384 & 0.61288 & 0.10160 & 0.047 & 5.18 & c  \\
OGLE-BLG-RRLYR-07508 & 0.29538 & 0.18069 & 0.61174 & 0.13520 & 0.018 & 4.64 & a,d,c  \\
OGLE-BLG-RRLYR-07518 & 0.29361 & 0.18030 & 0.61408 & 0.13652 & 0.033 & 6.49 & a,c  \\
OGLE-BLG-RRLYR-08084 & 0.31470 & 0.19288 & 0.61291 & 0.12491 & 0.041 & 5.91 &   \\
OGLE-BLG-RRLYR-08125 & 0.27681 & 0.16951 & 0.61236 & 0.10187 & 0.034 & 5.22 & a,e  \\
OGLE-BLG-RRLYR-08151 & 0.33179 & 0.20372 & 0.61399 & 0.12434 & 0.032 & 4.83 & a  \\
OGLE-BLG-RRLYR-08183 & 0.29386 & 0.18054 & 0.61439 & 0.13829 & 0.032 & 5.82 &   \\
OGLE-BLG-RRLYR-08214 & 0.31068 & 0.19062 & 0.61358 & 0.12252 & 0.027 & 4.77 &   \\
OGLE-BLG-RRLYR-08349 & 0.29709 & 0.18203 & 0.61272 & 0.12765 & 0.026 & 5.18 & a,c  \\
OGLE-BLG-RRLYR-08421 & 0.30188 & 0.18463 & 0.61160 & 0.12604 & 0.045 & 4.69 &   \\
OGLE-BLG-RRLYR-08475 & 0.31095 & 0.19070 & 0.61328 & 0.11270 & 0.034 & 4.77 &   \\
OGLE-BLG-RRLYR-08591 & 0.30220 & 0.18475 & 0.61133 & 0.12591 & 0.040 & 4.6 &   \\
OGLE-BLG-RRLYR-08674 & 0.31692 & 0.19426 & 0.61295 & 0.11828 & 0.028 & 4.93 & a  \\
OGLE-BLG-RRLYR-08721 & 0.24132 & 0.14779 & 0.61245 & 0.07841 & 0.047 & 4.84 & c  \\
OGLE-BLG-RRLYR-08745 & 0.29345 & 0.18031 & 0.61444 & 0.13072 & 0.025 & 5.02 & a,c  \\
OGLE-BLG-RRLYR-09134 & 0.29693 & 0.18243 & 0.61440 & 0.12569 & 0.023 & 5.28 & e  \\
OGLE-BLG-RRLYR-09164 & 0.28639 & 0.17557 & 0.61305 & 0.14166 & 0.022 & 4.53 & c  \\
OGLE-BLG-RRLYR-09206 & 0.29500 & 0.18069 & 0.61250 & 0.12988 & 0.054 & 4.79 &   \\
OGLE-BLG-RRLYR-09267 & 0.24988 & 0.15303 & 0.61242 & 0.12069 & 0.033 & 4.79 &   \\
OGLE-BLG-RRLYR-09305 & 0.30295 & 0.18573 & 0.61309 & 0.12764 & 0.037 & 5.04 & a,c  \\
OGLE-BLG-RRLYR-09349 & 0.24917 & 0.15275 & 0.61303 & 0.11188 & 0.023 & 6.2 & a  \\
OGLE-BLG-RRLYR-09529 & 0.30693 & 0.18840 & 0.61381 & 0.12600 & 0.019 & 4.7 & a,d  \\
OGLE-BLG-RRLYR-09733 & 0.31378 & 0.19227 & 0.61275 & 0.11894 & 0.028 & 4.83 & a  \\
OGLE-BLG-RRLYR-09929 & 0.29917 & 0.18347 & 0.61326 & 0.12792 & 0.026 & 5.5 &   \\
OGLE-BLG-RRLYR-10037 & 0.32980 & 0.20765 & 0.62963 & 0.12849 & 0.021 & 5.57 & s  \\
OGLE-BLG-RRLYR-10244 & 0.31346 & 0.19226 & 0.61333 & 0.12686 & 0.029 & 7.19 & a  \\
OGLE-BLG-RRLYR-10352 & 0.29857 & 0.18338 & 0.61419 & 0.12863 & 0.034 & 4.93 & a  \\
OGLE-BLG-RRLYR-10371 & 0.28934 & 0.17751 & 0.61347 & 0.13159 & 0.025 & 7.43 & a,c  \\
OGLE-BLG-RRLYR-10426 & 0.30200 & 0.18517 & 0.61314 & 0.12350 & 0.046 & 5.98 & c  \\
OGLE-BLG-RRLYR-10663 & 0.27700 & 0.16969 & 0.61259 & 0.08478 & 0.037 & 4.69 & f  \\
OGLE-BLG-RRLYR-10685 & 0.30007 & 0.18407 & 0.61341 & 0.13059 & 0.038 & 5.46 & a  \\
OGLE-BLG-RRLYR-10787 & 0.27651 & 0.16942 & 0.61272 & 0.11253 & 0.027 & 4.6 & a  \\
OGLE-BLG-RRLYR-11015 & 0.28958 & 0.17781 & 0.61403 & 0.12902 & 0.024 & 6.16 &   \\
OGLE-BLG-RRLYR-11107 & 0.30866 & 0.18957 & 0.61418 & 0.12575 & 0.030 & 5.79 & a  \\
OGLE-BLG-RRLYR-11151 & 0.29670 & 0.18211 & 0.61377 & 0.12594 & 0.030 & 5.51 & a,c  \\
OGLE-BLG-RRLYR-11503 & 0.31098 & 0.19028 & 0.61187 & 0.12873 & 0.027 & 6.01 & a,c  \\
OGLE-BLG-RRLYR-11559 & 0.30529 & 0.18630 & 0.61025 & 0.13199 & 0.023 & 4.53 & a  \\
OGLE-BLG-RRLYR-11641 & 0.32269 & 0.20344 & 0.63044 & 0.13423 & 0.022 & 5.03 & b,s  \\
OGLE-BLG-RRLYR-11684 & 0.30179 & 0.18530 & 0.61402 & 0.13118 & 0.029 & 5.52 & c  \\
OGLE-BLG-RRLYR-12037 & 0.31909 & 0.20128 & 0.63080 & 0.13177 & 0.025 & 4.56 & f,s  \\
OGLE-BLG-RRLYR-12071 & 0.33048 & 0.20777 & 0.62868 & 0.11846 & 0.021 & 4.69 & a,d  \\
OGLE-BLG-RRLYR-12099 & 0.29872 & 0.18351 & 0.61431 & 0.12941 & 0.028 & 5.79 & a  \\
OGLE-BLG-RRLYR-12113 & 0.31059 & 0.18936 & 0.60967 & 0.12928 & 0.022 & 5.91 & a  \\
\hline
\end{tabular}
\end{minipage}
\end{table*}

\begin{table*} 
 \centering
 \contcaption{}
 \begin{minipage}{140mm}
  \begin{tabular}{@{}lccccccc@{}}
  \hline
Name & $P_{\rm 1}$\thinspace[d] & $P_{\rm X}$\thinspace[d] & $P_{\rm X}/P_{\rm 1}$ & $A_{\rm 1}$\thinspace[mag] & $A_{\rm X}/A_{\rm 1}$ &  $S/N$ & Remarks            \\
 \hline
OGLE-BLG-RRLYR-12154 & 0.28683 & 0.17594 & 0.61341 & 0.13660 & 0.023 & 5.17 & a  \\
OGLE-BLG-RRLYR-12160 & 0.30123 & 0.18491 & 0.61387 & 0.13002 & 0.028 & 6.59 & a,c  \\
OGLE-BLG-RRLYR-12261 & 0.30507 & 0.18686 & 0.61252 & 0.12749 & 0.024 & 5.23 & a,c  \\
OGLE-BLG-RRLYR-12363 & 0.31900 & 0.19553 & 0.61294 & 0.12114 & 0.019 & 4.57 & a,d,c  \\
OGLE-BLG-RRLYR-12439 & 0.29558 & 0.18151 & 0.61409 & 0.13217 & 0.021 & 5.23 & c  \\
OGLE-BLG-RRLYR-12672 & 0.23540 & 0.14459 & 0.61423 & 0.09451 & 0.053 & 6.89 & c  \\
OGLE-BLG-RRLYR-12686 & 0.29624 & 0.18200 & 0.61436 & 0.12926 & 0.017 & 4.63 & a,g  \\
OGLE-BLG-RRLYR-12723 & 0.24919 & 0.15286 & 0.61342 & 0.12065 & 0.022 & 5.41 & c  \\
OGLE-BLG-RRLYR-13003 & 0.30888 & 0.18966 & 0.61402 & 0.11866 & 0.026 & 5.64 & a  \\
OGLE-BLG-RRLYR-13129 & 0.31144 & 0.19086 & 0.61284 & 0.11584 & 0.034 & 6.34 & c  \\
OGLE-BLG-RRLYR-13136 & 0.27940 & 0.17130 & 0.61310 & 0.11205 & 0.032 & 7.2 & b,c,g  \\
OGLE-BLG-RRLYR-13269 & 0.29716 & 0.18175 & 0.61162 & 0.12680 & 0.030 & 4.65 & f  \\
OGLE-BLG-RRLYR-13331 & 0.29378 & 0.18047 & 0.61433 & 0.12848 & 0.051 & 5.29 &   \\
OGLE-BLG-RRLYR-13401 & 0.30025 & 0.18374 & 0.61198 & 0.13070 & 0.029 & 5.05 & c  \\
OGLE-BLG-RRLYR-13422 & 0.30212 & 0.18519 & 0.61297 & 0.13597 & 0.028 & 6.66 & a,c  \\
OGLE-BLG-RRLYR-13550 & 0.32117 & 0.19672 & 0.61251 & 0.12606 & 0.025 & 4.77 & g  \\
OGLE-BLG-RRLYR-14404 & 0.24538 & 0.15020 & 0.61209 & 0.09020 & 0.048 & 5.35 & b,c  \\
OGLE-BLG-RRLYR-14502 & 0.29471 & 0.18080 & 0.61350 & 0.12590 & 0.027 & 4.95 & a  \\
OGLE-BLG-RRLYR-14677 & 0.29520 & 0.18135 & 0.61432 & 0.12276 & 0.033 & 5.5 & a  \\
OGLE-BLG-RRLYR-14687 & 0.27104 & 0.16629 & 0.61352 & 0.11493 & 0.047 & 5.59 &   \\
OGLE-BLG-RRLYR-14744 & 0.29256 & 0.17932 & 0.61291 & 0.13485 & 0.022 & 4.57 & a  \\
OGLE-BLG-RRLYR-15271 & 0.28991 & 0.17771 & 0.61299 & 0.14640 & 0.023 & 5.03 &   \\
OGLE-BLG-RRLYR-16157 & 0.29843 & 0.18333 & 0.61431 & 0.11866 & 0.060 & 5.75 & a  \\ 
\hline
\multicolumn{8}{l}{a -- period change; b -- suspected Blazhko effect; c -- additional mode is non-stationary; }\\
\multicolumn{8}{l}{d -- time-dependent prewhitening was used; e -- additional signal present in the data; f -- daily alias is higher;}\\
\multicolumn{8}{l}{g -- combination frequencies detected; s -- sub-harmonic of additional mode detected}\\
\hline
\end{tabular}
\end{minipage}
\end{table*}

\begin{table*} 
 \centering
 \begin{minipage}{140mm}
 \caption{Same as Tab.~\ref{table:pewne}, but for star with non-radial mode detected with $S/N < 4.5$.}
 \label{table:kandydatki}
  \begin{tabular}{@{}lccccccc@{}}
  \hline
Name & $P_{\rm 1}$\thinspace[d] & $P_{\rm X}$\thinspace[d] & $P_{\rm X}/P_{\rm 1}$ & $A_{\rm 1}$\thinspace[mag] & $A_{\rm X}/A_{\rm 1}$ &  $S/N$ & Remarks            \\
 \hline
OGLE-BLG-RRLYR-00679 & 0.30903 & 0.18961 & 0.61356 & 0.11982 & 0.055 & 4.47 &   \\
OGLE-BLG-RRLYR-01744 & 0.29585 & 0.18111 & 0.61216 & 0.13412 & 0.030 & 4.47 &   \\
OGLE-BLG-RRLYR-02027 & 0.27862 & 0.17015 & 0.61069 & 0.11673 & 0.14 & 4.22 & RRd  \\ 
OGLE-BLG-RRLYR-02077 & 0.29234 & 0.17953 & 0.61409 & 0.12724 & 0.033 & 4.37 &   \\
OGLE-BLG-RRLYR-02478 & 0.39036 & 0.23872 & 0.61154 & 0.11254 & 0.054 & 4.14 & a,d  \\
OGLE-BLG-RRLYR-02615 & 0.29403 & 0.18024 & 0.61301 & 0.13983 & 0.026 & 4.16 & a  \\
OGLE-BLG-RRLYR-04466 & 0.29726 & 0.18212 & 0.61266 & 0.12641 & 0.033 & 4.37 &   \\
OGLE-BLG-RRLYR-04754 & 0.28631 & 0.17577 & 0.61392 & 0.12870 & 0.036 & 4.39 &   \\
OGLE-BLG-RRLYR-05550 & 0.24910 & 0.15332 & 0.61548 & 0.13672 & 0.019 & 4.34 & a,e  \\
OGLE-BLG-RRLYR-05956 & 0.29174 & 0.17951 & 0.61529 & 0.13119 & 0.034 & 4.01 &   \\
OGLE-BLG-RRLYR-07027 & 0.30650 & 0.18807 & 0.61361 & 0.12118 & 0.028 & 4.12 &   \\
OGLE-BLG-RRLYR-07076 & 0.29638 & 0.18094 & 0.61048 & 0.12487 & 0.022 & 4.02 & s  \\
OGLE-BLG-RRLYR-07368 & 0.31903 & 0.19547 & 0.61272 & 0.11788 & 0.025 & 4.04 &   \\
OGLE-BLG-RRLYR-07980 & 0.29529 & 0.18033 & 0.61069 & 0.12154 & 0.016 & 4.33 & a  \\
OGLE-BLG-RRLYR-08028 & 0.28384 & 0.17423 & 0.61383 & 0.13588 & 0.024 & 4.48 & a  \\
OGLE-BLG-RRLYR-08048 & 0.31430 & 0.19243 & 0.61225 & 0.13109 & 0.020 & 4.32 & a  \\
OGLE-BLG-RRLYR-08321 & 0.33110 & 0.20317 & 0.61361 & 0.13587 & 0.037 & 4.42 & a  \\
OGLE-BLG-RRLYR-08396 & 0.29395 & 0.18045 & 0.61389 & 0.13314 & 0.019 & 4.49 &   \\
OGLE-BLG-RRLYR-08443 & 0.30694 & 0.18815 & 0.61301 & 0.15399 & 0.014 & 4.29 &   \\
OGLE-BLG-RRLYR-08597 & 0.32093 & 0.20236 & 0.63053 & 0.13686 & 0.028 & 4.2 & a  \\
OGLE-BLG-RRLYR-08640 & 0.28460 & 0.17460 & 0.61349 & 0.12978 & 0.032 & 4.36 & a  \\
OGLE-BLG-RRLYR-08716 & 0.28834 & 0.17709 & 0.61415 & 0.13928 & 0.017 & 4.33 & a,d    \\
OGLE-BLG-RRLYR-08788 & 0.28739 & 0.17632 & 0.61353 & 0.13720 & 0.029 & 4.19 & e  \\
OGLE-BLG-RRLYR-08818 & 0.31261 & 0.19193 & 0.61396 & 0.12928 & 0.023 & 4.38 & a,d  \\
OGLE-BLG-RRLYR-08896 & 0.28189 & 0.17307 & 0.61394 & 0.14593 & 0.020 & 4.47 & f  \\
OGLE-BLG-RRLYR-09481 & 0.31772 & 0.19466 & 0.61267 & 0.12172 & 0.041 & 4.18 &   \\
OGLE-BLG-RRLYR-09620 & 0.29163 & 0.17887 & 0.61334 & 0.13539 & 0.038 & 4.16 &   \\
OGLE-BLG-RRLYR-10219 & 0.29384 & 0.17993 & 0.61233 & 0.13964 & 0.019 & 4.24 &   \\
OGLE-BLG-RRLYR-10230 & 0.31078 & 0.19069 & 0.61358 & 0.12161 & 0.034 & 4.33 & a  \\
OGLE-BLG-RRLYR-10392 & 0.24760 & 0.15193 & 0.61361 & 0.11625 & 0.030 & 4.31 & c  \\
OGLE-BLG-RRLYR-10756 & 0.29659 & 0.18232 & 0.61470 & 0.14130 & 0.027 & 4.47 & a  \\
OGLE-BLG-RRLYR-10880 & 0.24889 & 0.15776 & 0.63384 & 0.10672 & 0.049 & 4.24 &   \\
OGLE-BLG-RRLYR-10951 & 0.30553 & 0.18685 & 0.61157 & 0.13935 & 0.017 & 4.02 &   \\
OGLE-BLG-RRLYR-11063 & 0.28947 & 0.17783 & 0.61431 & 0.13518 & 0.021 & 4.43 & f  \\
OGLE-BLG-RRLYR-11575 & 0.29661 & 0.18172 & 0.61264 & 0.12730 & 0.035 & 4.32 & f  \\
OGLE-BLG-RRLYR-11711 & 0.30466 & 0.18712 & 0.61420 & 0.12189 & 0.028 & 4.12 & a,d  \\
OGLE-BLG-RRLYR-11726 & 0.30195 & 0.18538 & 0.61396 & 0.12498 & 0.049 & 4.03 & e  \\
OGLE-BLG-RRLYR-11742 & 0.27929 & 0.17185 & 0.61532 & 0.13836 & 0.017 & 4.36 & a  \\
OGLE-BLG-RRLYR-11761 & 0.30881 & 0.18939 & 0.61329 & 0.12562 & 0.027 & 4.03 &   \\
OGLE-BLG-RRLYR-11789 & 0.26796 & 0.16397 & 0.61190 & 0.11555 & 0.030 & 4.46 &   \\
OGLE-BLG-RRLYR-11806 & 0.31314 & 0.19207 & 0.61337 & 0.11690 & 0.053 & 4.21 &   \\
OGLE-BLG-RRLYR-11917 & 0.31333 & 0.19193 & 0.61255 & 0.12334 & 0.020 & 4.01 & c  \\
OGLE-BLG-RRLYR-11942 & 0.22411 & 0.13788 & 0.61525 & 0.07636 & 0.047 & 4.2 & f  \\
OGLE-BLG-RRLYR-12127 & 0.28870 & 0.17740 & 0.61445 & 0.14603 & 0.016 & 4.3 & a,d  \\
OGLE-BLG-RRLYR-12315 & 0.28681 & 0.17545 & 0.61171 & 0.14384 & 0.012 & 4.1 & a,d,f  \\
OGLE-BLG-RRLYR-12355 & 0.35181 & 0.21630 & 0.61481 & 0.14052 & 0.081 & 4.15 &   \\
OGLE-BLG-RRLYR-12369 & 0.26545 & 0.16311 & 0.61445 & 0.11163 & 0.031 & 4.18 &   \\
OGLE-BLG-RRLYR-12421 & 0.33805 & 0.21317 & 0.63059 & 0.12349 & 0.024 & 4.17 & a  \\
OGLE-BLG-RRLYR-12692 & 0.32502 & 0.19672 & 0.60527 & 0.10872 & 0.059 & 4.19 & a,d  \\
OGLE-BLG-RRLYR-13099 & 0.28072 & 0.17270 & 0.61520 & 0.14614 & 0.014 & 4.12 & a  \\
OGLE-BLG-RRLYR-13184 & 0.31028 & 0.18985 & 0.61186 & 0.12736 & 0.040 & 4.32 &   \\
OGLE-BLG-RRLYR-13247 & 0.29538 & 0.18147 & 0.61437 & 0.12987 & 0.060 & 4.08 &   \\
OGLE-BLG-RRLYR-13340 & 0.29206 & 0.17893 & 0.61263 & 0.13633 & 0.038 & 4.25 & f  \\
OGLE-BLG-RRLYR-13528 & 0.29747 & 0.18258 & 0.61376 & 0.14883 & 0.027 & 4.1 &   \\
OGLE-BLG-RRLYR-13911 & 0.24996 & 0.15302 & 0.61220 & 0.12379 & 0.028 & 4.38 &   \\
OGLE-BLG-RRLYR-14031 & 0.42978 & 0.26470 & 0.61592 & 0.11555 & 0.031 & 4.13 & RRd,a \\
OGLE-BLG-RRLYR-14162 & 0.28050 & 0.16693 & 0.59512 & 0.08493 & 0.020 & 4.25 & a,d  \\
OGLE-BLG-RRLYR-14190 & 0.31324 & 0.19199 & 0.61291 & 0.12020 & 0.040 & 4.4 & a,f  \\
OGLE-BLG-RRLYR-14344 & 0.30933 & 0.18979 & 0.61355 & 0.12268 & 0.030 & 4.08 & g  \\
OGLE-BLG-RRLYR-14467 & 0.26591 & 0.16291 & 0.61264 & 0.11933 & 0.026 & 4.21 & c,f  \\
OGLE-BLG-RRLYR-14731 & 0.33149 & 0.20292 & 0.61215 & 0.14174 & 0.022 & 4.48 & a  \\
OGLE-BLG-RRLYR-14917 & 0.30481 & 0.18748 & 0.61507 & 0.12575 & 0.027 & 4.23 &   \\
 & 0.30481 & 0.18637 & 0.61143 & 0.12539 & 0.031 & 4.61 &   \\
\hline
\end{tabular}
\end{minipage}
\end{table*}

\begin{table*} 
 \centering
 \contcaption{}
 \begin{minipage}{140mm}
  \begin{tabular}{@{}lccccccc@{}}
  \hline
Name & $P_{\rm 1}$\thinspace[d] & $P_{\rm X}$\thinspace[d] & $P_{\rm X}/P_{\rm 1}$ & $A_{\rm 1}$\thinspace[mag] & $A_{\rm X}/A_{\rm 1}$ &  $S/N$ & Remarks            \\
 \hline
OGLE-BLG-RRLYR-15672 & 0.25924 & 0.15874 & 0.61230 & 0.11762 & 0.048 & 4.08 &   \\
OGLE-BLG-RRLYR-16265 & 0.31311 & 0.19745 & 0.63059 & 0.12817 & 0.052 & 4.02 &   \\

\hline
\multicolumn{8}{l}{a -- period change; b -- suspected Blazhko effect; c -- additional mode is non-stationary; }\\
\multicolumn{8}{l}{d -- time-dependent prewhitening was used; e -- additional signal present in the data; f -- daily alias is higher;}\\
\multicolumn{8}{l}{g -- combination frequencies detected; s -- sub-harmonic of additional mode detected}\\
\hline
\end{tabular}
\end{minipage}
\end{table*}

\section{Discussion}\label{sec.discussion}

We have significantly increased a sample of RRc stars with additional non-radial mode which has characteristic period ratio about 0.61 with the first overtone. Eighteen stars were known before, now we detected 147 additional. In Fig. \ref{fig.new} we show how newly discovered stars are distributed in the Petersen diagram. 137 stars are placed between period ratios 0.605--0.616, with average value 0.613, and form a well defined horizontal sequence. We found no significant correlation between period ratio and the first overtone period. Nine stars, with both low and high $S/N$ detections, have period ratios about 0.63 and they form a separate, second group on the Petersen diagram. For these stars minimum, maximum and average values are 0.62868, 0.63384 and 0.631, respectively. This sequence seems to be well defined. There is a clear gap between the two sequences. Within the second group we observe a slight trend of increasing period ratio with the decreasing period of the first overtone. We note that in the case of first overtone Cepheids with additional mode three well separated sequences are observed  \citep[see][]{ogle_rr_smc}. In addition to the two sequences just described one star has significantly different period ratio, below 0.6, but it is a weak detection.

\begin{figure}
\centering
\resizebox{\hsize}{!}{\includegraphics{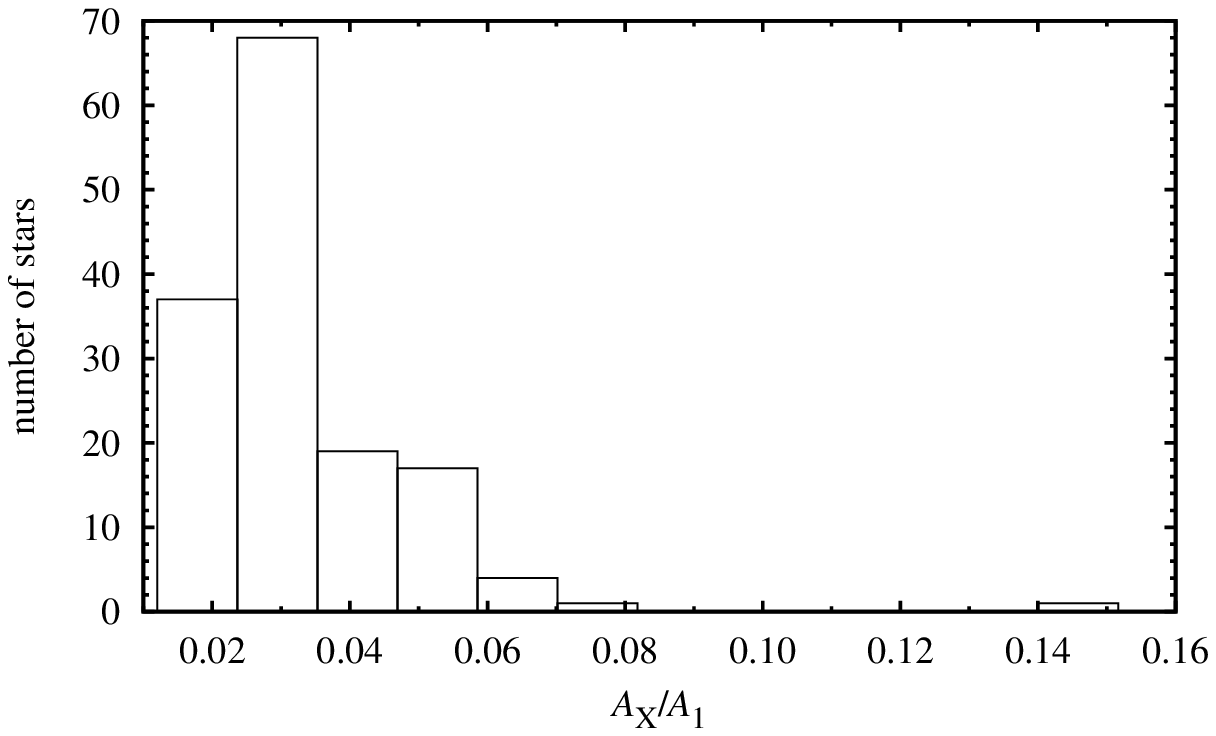}}
\caption{The histogram of amplitude ratios for the newly discovered stars.}
\label{fig.ar}
\end{figure}

For all stars the amplitude of the additional mode is very low compared to the amplitude of the first overtone (see Tabs.~\ref{table:pewne} and \ref{table:kandydatki}). Amplitude of the additional mode amounts to $1.2-8.1$ per cent of the first overtone amplitude, with modal value $2.8$ per cent. The highest amplitude ratio is detected for RRd star, where $A_{\rm X}/A_1$ is $14$ per cent. Fig.~\ref{fig.ar} shows the histogram of amplitude ratios for all detected stars.

\begin{figure}
\centering
\resizebox{\hsize}{!}{\includegraphics{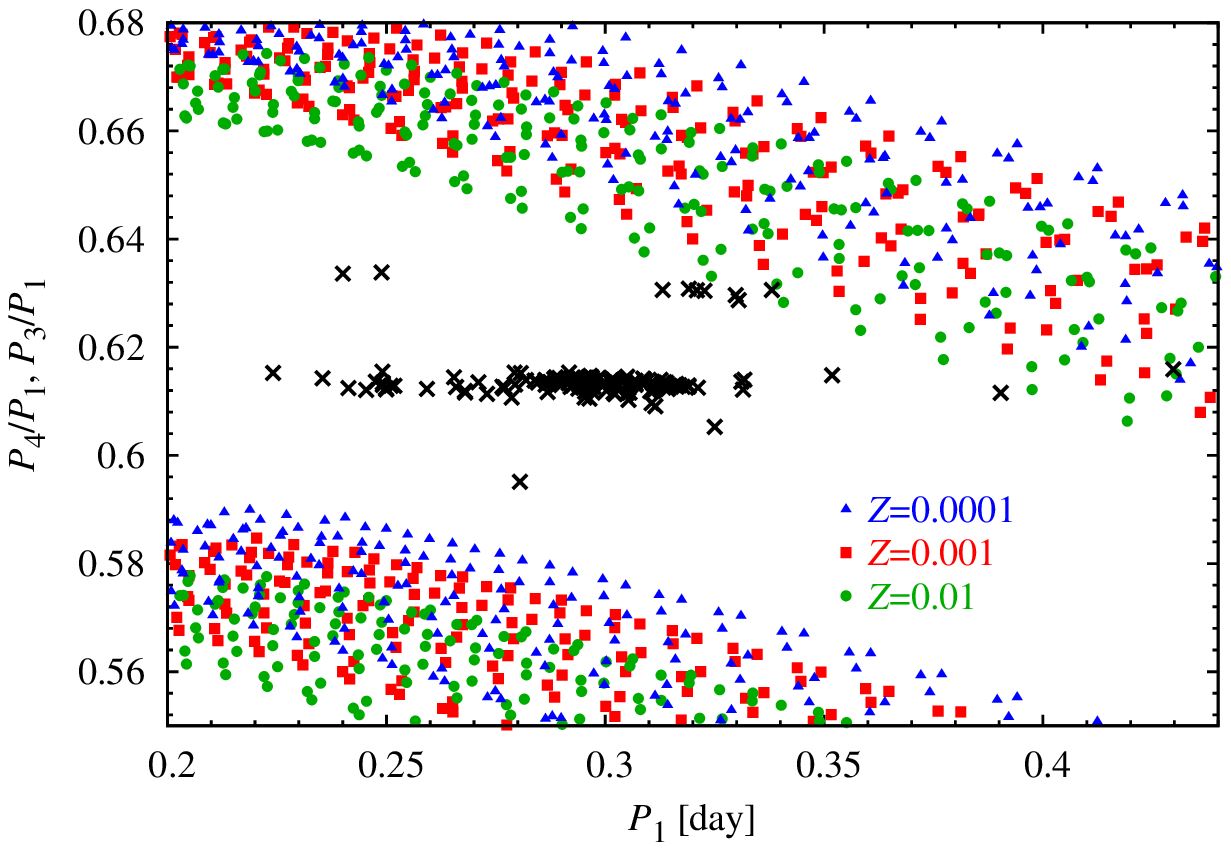}}
\caption{The Petersen diagram for newly detected double-mode RR~Lyrae stars confronted with predicted $P_3/P_1$ and $P_4/P_1$ radial mode period ratios computed with the Warsaw codes.}
\label{fig.pet_models}
\end{figure}

The identification of additional mode as non-radial results from comparison with pulsation models which, for given model parameters (mass, $M$, luminosity, $L$, effective temperature, $T_{\rm eff}$, and chemical composition, $X/Z$), provide periods of the radial modes and their linear growth rates. This computations indicate that period ratios observed both in the first overtone Cepheids and in RRc stars with additional $\sim 0.6$ mode cannot correspond to period ratios between radial modes -- see \cite{wdrs09} \& \cite{wd12} for Cepheid models and \cite{pamsm14}, for RR~Lyrae models. They are in between the expected third-to-first overtone period ratio and fourth-to-first overtone period ratio. In Fig.~\ref{fig.pet_models} we present a comparison between model and observed period ratios for our sample of RRc stars. The models were computed with the Warsaw pulsation codes \citep{sm08} assuming different values for $M$ ($0.55-0.75{\rm M}_\odot$), $L$ ($40-70{\rm L}_\odot$), $T_{\rm eff}$ (across full instability strip for each $M/L$) and metallicity ($0.01$, $0.001$ and $0.0001$; different symbols/colors in Fig.~\ref{fig.pet_models}). No doubt, the main $0.613$ sequence cannot correspond to double-mode radial pulsation. It falls exactly in between $P_3/P_1$ and $P_4/P_1$ model sequences. Only two stars from its long-period tail fall among the $P_3/P_1$ model period ratios for metal rich stars. The situation is less clear for the $0.631$ sequence, which fits the lower boundary of $P_3/P_1$ model ratios. We argue however, that also these stars are radial--non-radial double-mode stars. First, observed period ratios are lower than the model period ratios and form a group, in which increase of period ratio with decreasing period is only very slight, while in the models the increase is well visible. Second, the models that directly neighbour the discussed group correspond to high metallicities, $Z=0.01$, which is not typical for RR~Lyrae stars. Third, such form of pulsation, i.e., 1O+3O double-mode pulsation, although detected previously in two first overtone Cepheids \citep{ogle_freaks}, is certainly difficult to explain. Models predict that third overtone is linearly stable. At the moment we do not have an explanation how such form of pulsation could arise, we admit however, that the problem of the origin of double-mode pulsation of any flavour, even in the most frequent F+1O RRd stars or double-mode Cepheids is difficult and unsolved \citep[see][]{sm10, smolec14}.

\begin{figure}
\centering
\resizebox{\hsize}{!}{\includegraphics{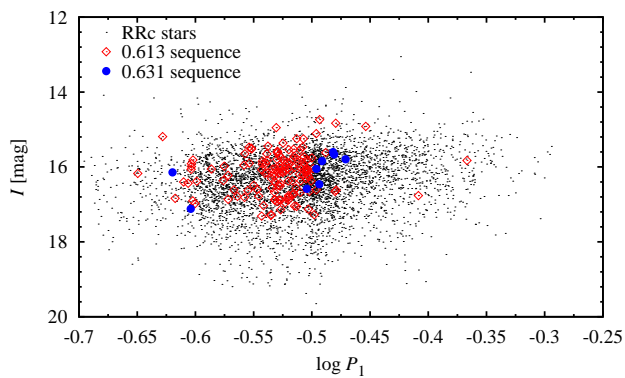}}
\caption{The period-luminosity diagram for all RRc stars from the Galactic bulge. Stars with the additional mode are marked with different symbols.}
\label{fig.p_mag}
\end{figure}

\begin{figure}
\centering
\resizebox{\hsize}{!}{\includegraphics{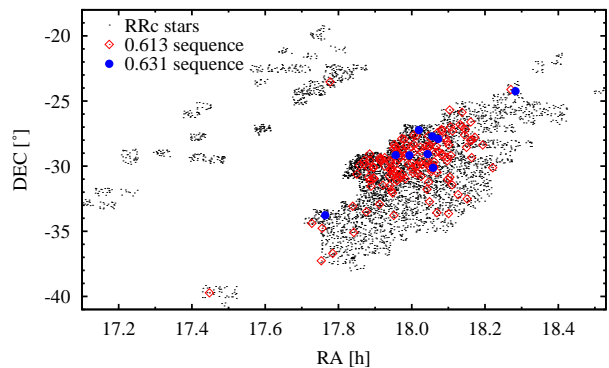}}
\caption{Location in the sky of the newly discovered stars. All RRc stars from the OGLEIII Galactic bulge catalog are plotted with small dots.}
\label{fig.pol}
\end{figure}

To check whether stars with the additional mode differ from other RRc stars we have plotted the location of OGLE Galactic bulge RRc stars in the period-luminosity diagram in Fig. \ref{fig.p_mag}. In addition we have plotted the stars from the two sequences (0.613 and 0.631) with different symbols. Stars with the additional mode occupy a shorter period part of this diagram, but do not group at any specific place. Except two, all stars from 0.631 sequence have longer periods than majority of stars from 0.613 sequence, but otherwise they cover a similar luminosity range. To check the possibility that stars with higher period ratio belong to other stellar system that coincide with the Galactic bulge (e.g. Sagittarius Dwarf Galaxy, cluster of stars) we analysed their location in the sky -- Fig.~\ref{fig.pol}. The stars group in the fields close to the Galactic centre, which is expected as these are the most dense stellar fields. No grouping corresponding to any cluster is visible on Fig. \ref{fig.pol}.

\begin{figure}
\centering
\resizebox{\hsize}{!}{\includegraphics{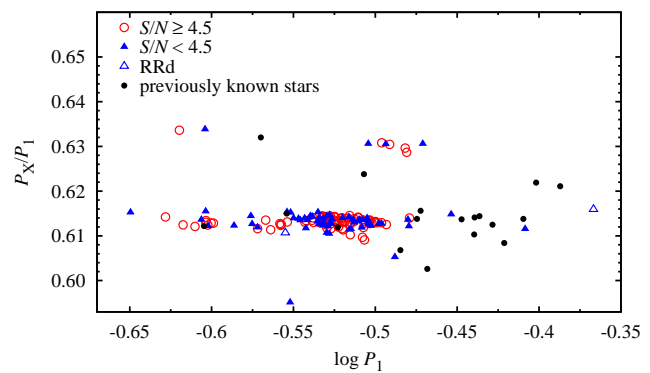}}
\caption{The Petersen diagram for newly discovered and already known stars with the additional periodicity.}
\label{fig.newold}
\end{figure}

In order to compare new stars with the previously known, we plotted them together on the Petersen diagram in Fig.~\ref{fig.newold}. New stars have shorter periods than those already known, so these two groups barely overlap. Some of the previously known stars nicely fit the 0.613 sequence defined by the newly discovered stars or its long-period extension. There are stars with higher period ratios, but except one, they do not fit well to 0.631 sequence. Stars known so far have large dispersion on the diagram, much larger than the Galactic bulge sample. We suppose that this situation is caused by a population effects, as in the old sample we have stars from different stellar systems ($\omega$ Centauri, LMC, $Kepler$ field).

\begin{figure}
\centering
\resizebox{\hsize}{!}{\includegraphics{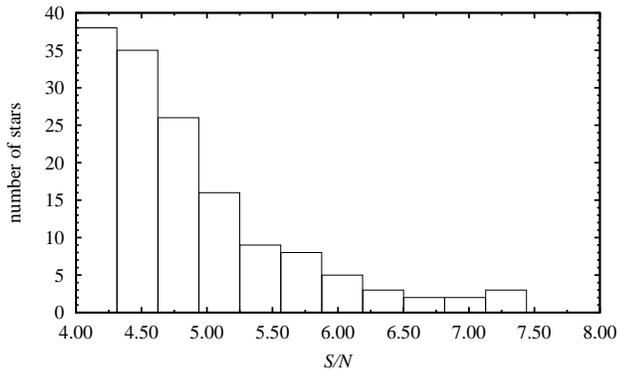}}
\caption{The histogram of signal-to-noise ratio for the peaks corresponding to non-radial mode.}
\label{fig.sn}
\end{figure}

In all RRc and RRd stars observed by the space telescopes we see additional mode \citep{aqleo,pam14,szabo_corot}, which indicates that this mode should be common among these stars. In all cases the amplitude of the additional mode is small, in the mmag regime. The high precision of the space photometry made its detection possible. The ground based observations have much higher noise level. We can expect that only stars with the highest intrinsic amplitude of the additional mode, which are likely a small fraction of this presumably large group, could be detected from the ground. In Fig. \ref{fig.sn} we show the histogram of $S/N$ of our detections that seem to support such view. The number of stars grows significantly as $S/N$ is decreased till a cutoff value of 4 resulting from our arbitrarily chosen criterion. Clearly, there must be much more RRc stars in the Galactic bulge with the additional non-radial mode, but because of its low amplitude, they are hidden in the noise. 147 stars detected in this study are likely just the tip of the iceberg.

Among the previously known stars with additional mode none shows the Blazhko modulation. From newly discovered sample we selected four possible candidates for Blazhko stars (marked with `b' in `remarks' column of Tabs.~\ref{table:pewne} and \ref{table:kandydatki}). Three stars have additional close frequency at $f_1$ (dublets) which may correspond to modulation periods of 1115, 308 and 12 days. One star has triplet at the primary frequency, but the sidepeaks are not well resolved at the noise level observed in this star. In addition frequency spectrum at $f_1$ is rather complex with non-stationary peaks. We did not find a star with typical Blazhko modulation which shows additional non-radial mode, but we cannot exclude that such stars exist. Proper analysis of stars with the Blazhko effect will be the subject of future research.

\begin{figure}
\centering
\resizebox{\hsize}{!}{\includegraphics{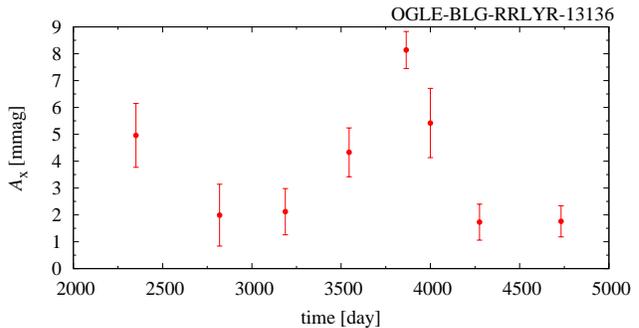}}
\caption{The seasonal amplitude change of the additional non-radial mode in OGLE-BLG-RRLYR-13136.}
\label{fig.ac}
\end{figure}

In several stars we detect unresolved signal at $f_{\rm X}$ or signal at $f _{\rm X}$ is accompanied by additional close peaks that may also be non-stationary. This situation is common to RRc stars observed from space, both by {\it CoRoT} \citep{szabo_corot} and by {\it Kepler} \citep{pamsm14}. The analysis of nearly continuous and top-quality {\it Kepler} photometry indicates that it corresponds to quasi-periodic modulation of $f_{\rm X}$. Likely we detect a signature of the same phenomenon. In Fig.~\ref{fig.ac} we show the seasonal variation of $A_{\rm X}$ for one star from our sample. Results are very similar for other stars -- the amplitude variation is sometimes regular, sometimes more erratic. We note that with poorly sampled and noisy ground-based data we cannot resolve the variations that occur on a time scale shorter than the season length.

Finally, in four stars, of which three are members of the 0.631 group in the Petersen diagram, we detect additional significant signal at $\sim 1/2f_{\rm X}$, which we identify as sub-harmonic of $f_{\rm X}$. Tab.~\ref{tab:sh} provides basic information about this additional frequency, denoted in the following by $f_{\rm s}$. The presence of sub-harmonic may indicate that non-radial mode undergoes period doubling \citep[see e.g.][]{blher}. We do not detect other sub-harmonics, e.g. at $3/2f_{\rm X}$. Sub-harmonics, at $\sim 1/2f_{\rm X}$ and at $\sim 3/2f_{\rm X}$ are often detected in space data for RRc stars \citep{pamsm14}, both in stars with period ratios close to 0.61 and in stars with larger period ratios. Just as we found here, they may be located slightly off the exact sub-harmonic of $f_{\rm X}$. Amplitude of the sub-harmonics we detect is a significant fraction of non-radial's mode amplitude, $A_{\rm s}/A_{\rm X}$ is larger than 0.8 for all stars and in OGLE-BLG-RRLYR-12037 the amplitudes are equal. In addition, in OGLE-BLG-RRLYR-10037 and in OGLE-BLG-RRLYR-11641 sub-harmonic is non-stationary. In case of the {\it Kepler} data sub-harmonics are always non-stationary. Both $A_{\rm s}$ and $A_{\rm X}$ vary in time. Although typically $A_{\rm s}$ is lower than $A_{\rm X}$ the reverse situation may also happen. We also note that sub-harmonic is likely present in other star, OGLE-BLG-RRLYR-08674, but $S/N$ is below 4 for the suspected peak.  

\begin{table*}
\centering
\caption{Stars with additional non-radial mode and additional signal close to its sub-harmonic.}
\label{tab:sh}
\begin{tabular}{lrrrrrr}
\hline
star & $P_{\rm X}/P_1$ & $f_x$ & $f_{\rm s}$ & $S/N$ & $f_{\rm s}/f_{\rm X}$ & $A_{\rm s}/A_{\rm X}$\\
\hline
OGLE-BLG-RRLYR-07076 & 0.61048 & 5.52684(2) &2.69643(2) & 4.11 & 0.488 & 0.93 \\ 
OGLE-BLG-RRLYR-10037 & 0.62963 & 4.81580(2) & 2.41252(2) & 4.76 & 0.501 & 0.92 \\ 
OGLE-BLG-RRLYR-11641 & 0.63044 & 4.91547(5) & 2.44944(2) & 4.52 & 0.498  & 0.86\\ 
OGLE-BLG-RRLYR-12037 & 0.63080 & 4.96815(4) & 2.47915(4) & 4.77 & 0.499 & 1.01 \\ 
\hline
\end{tabular}
\end{table*}

\section{Summary and conclusions}\label{sec.summary}

We have analysed 4989 stars from the Galactic bulge classified in OIII-CVS as RRc stars, and 91 stars classified as RRd. We searched for non-radial mode, which forms characteristic period ratio with the first overtone mode around 0.6. We detected 145 stars pulsating in two modes simultaneously: in the first overtone and in the non-radial mode. Two stars pulsate in the fundamental mode, in the first overtone (RRd) and in the non-radial mode.

The newly discovered stars constitute 3 per cent of the analysed sample. They form two sequences on the Petersen diagram. The sequence at the period ratio around 0.613 is tight, well defined and contains majority of found stars. Several stars form the second, less pronounced sequence at period ratio around 0.631.

Our research significantly increased the number of known stars showing the new form of double-mode pulsation (until now only 18 such stars were known). Models leave no doubt, the additional mode cannot correspond to a radial mode, it must be non-radial. The new group of double-mode radial--non-radial RR~Lyrae stars, postulated by \cite{pamsm14}, has now more than 160 members. Analysis of space photometry and our results suggest that this new form of pulsation must be common among RRc stars. Still, we do not understand mechanism behind this phenomenon: which mode is excited and how \citep[see however][]{wd12}. 

In many cases we find that the additional mode is non-stationary or has close companions in the frequency spectrum, which may be non-stationary as well. Similar behaviour was found in the {\it Kepler} and {\it CoRoT} data for RRc stars \citep{pamsm14,szabo_corot}. In four stars we detected sub-harmonics of the additional mode, which are often visible in the space photometry. They are a signature of period doubling of the non-radial mode. 

Although ground-based photometry is of much lower quality than space-borne observations, it has one crucial advantage -- it was collected for thousands of RR~Lyrae stars, making it possible to detect more than 100 objects of the new class. Future work will be focused on careful analysis of stars with Blazhko effect in search for this non-radial mode. We are also going to conduct the same analysis on stars from LMC, SMC and from clusters for which we have good ground-based photometric data.

In the Appendix we included a list of interesting stars found as a side effect of our study. The list contains five stars in which we found additional periodicity which may correspond to other radial mode (two HADS stars, two weak 1O+3O candidates), and 14 stars in which we found frequencies which do not correspond to any radial mode. For further details and brief analysis we refer the reader to the Appendix.

\section*{Acknowledgments}

This research is supported by the Polish National Science Centre through grant DEC-2012/05/B/ST9/03932. We acknowledge the summer student program at Nicolaus Copernicus Astronomical Center during which part of this work was completed.

\appendix

\section[]{Notes on individual RRc stars}

Below we briefly discuss the interesting stars found during our analysis. In most of them we detect additional significant signal in the frequency spectrum prewhitened with the first overtone and its harmonics, and possibly additional non-radial mode with $P_{\rm X}/P_{1}\in(0.58-0.64)$. In a few cases the period ratio fits the expected value for radial modes of either RR~Lyrae stars or High Amplitude $\delta$ Scuti Stars (HADS). These stars are discussed first. Then we discuss the stars for which period ratio does not fit the radial mode scenario. Below, $P_1$ always refer to primary period (first overtone), $P_{\rm X}$ to period of the non-radial mode of interest in this paper, and $P_{\rm a}$, $P_{\rm b}$,$\dots$, to additional periodicities detected in the data. Abbreviation ``no comb.'' means that no combination frequencies were found involving frequency of the additional signal and other frequencies present in the data. In such a case we do not have a proof that the frequencies originate from the same star. The term ``no cont.'' means we could not find a contamination source of similar period within 1 arc minute in the OIII-CVS. It does not exclude blending, as additional star may be too close to be resolved in the OGLE photometry, or was not included in OIII-CVS. We note that the list of stars with additional significant signal presented below is by no means complete. These are stars found only because they have additional $\sim 0.61$ mode as well, and so they where analysed in detail, or because the daily alias of additional signal falls into the frequency range in which we searched for $\sim 0.61$ mode.

\subsection{Stars with additional, likely radial modes}

\noindent {\bf OGLE-BLG-RRLYR-04259}. $P_1=0.2761459(2)$\thinspace d. In addition we find $P_{\rm a}=0.3639184(7)$\thinspace d ($S/N=13.3$). Both $P_1$ and $P_{\rm a}$ are non-stationary. Period ratio $P_1/P_{\rm a}=0.759$ suggests that additional signal corresponds to fundamental mode in HADS star of particularly long period \citep[for similar HADS stars see][]{poleski, pietruk}. No comb., no cont. Most likely F+1O HADS star.

\noindent {\bf OGLE-BLG-RRLYR-07028}.  $P_1=0.2441616(3)$\thinspace d. In addition we find $P_{\rm a}=0.1671389(3)$\thinspace d ($S/N=10.5$). Period ratio $(P_{\rm a}/P_1=0.685)$ indicates a possible 1O+3O double-mode star. For this star however, we do not see the harmonics of 1O, neither its combinations with possible 3O. No cont. Blend or weak 1O+3O double-mode RR~Lyrae candidate. 

\noindent {\bf OGLE-BLG-RRLYR-08788}. $P_1=0.28739168(6)$\thinspace d, $P_{\rm X}=0.176322(1)$\thinspace d. In addition we find a weak signal at  $P_{\rm a}=0.190487(2)$\thinspace d ($S/N=4.3$). $P_{\rm a}/P_1=0.663$ indicate a possible 1O+3O pulsator. It is also possible that its lower frequency alias is a true signal. No comb., no cont. Weak candidate for 1O+3O double-mode RR~Lyrae pulsator.

\noindent {\bf OGLE-BLG-RRLYR-14402}. $P_1=0.22033494(5)$\thinspace d. In addition we find $P_{\rm a}=0.16785954(4)$\thinspace d ($S/N=21.5$). Period ratio $P_{\rm a}/P_1=0.762$ perfectly fits the F+1O HADS progression in the Petersen diagram. Combination frequencies are present and strong ($f_1+f_{\rm a}, f_{\rm a}-f_1, 2f_{\rm a}+f_1$). Thus, we interpret $P_1$ and $P_{\rm a}$ as fundamental and first overtone modes, respectively,  in a firm HADS star.

\noindent {\bf OGLE-BLG-RRLYR-16581}. $P_1=0.2261119(2)$\thinspace d. In addition we find $P_{\rm a}=0.178914(2)$\thinspace d ($S/N=7.2$). The period ratio is $P_{\rm a}/P_1=0.791$ and so the star falls in between F+1O HADS stars and 1O+2O Cepheids. It may be a 1O+2O RR~Lyrae star. We detect only one harmonic of $f_1$. No comb., no cont. Likely a double-mode radial pulsator but exact identification is uncertain.

\subsection{Stars with additional, unidentified frequencies}

\noindent {\bf OGLE-BLG-RRLYR-01443}. $P_1=0.2392456(6)$\thinspace d. In addition we find $P_{\rm a}=0.1695492(9)$\thinspace d ($S/N=7.6$). Period ratio ($P_{\rm a}/P_1=0.709$) does not fit the radial mode scenario. No comb., no cont. Non-radial mode or blend.

\noindent {\bf OGLE-BLG-RRLYR-02251}. This star has two designations in OIII-CVS: RRc star OGLE-BLG-RRLYR-02251 and double-overtone 1O+2O Cepheid OGLE-BLG-CEP-04. We find $P_1=0.2400460(2)$\thinspace d, $P_2=0.1902512(6)$\thinspace d and so $P_2/P_1=0.793$ in principle fits the 1O/2O period ratio. We note however that 1O period is very short then, and in fact star is located at the intersection of 1O+2O Cepheids and 1O+2O HADS stars. In addition we detect $P_{\rm a}=0.168464(1)$\thinspace d ($S/N=4.8$, $P_{\rm a}/P_1=0.702$) and $P_{\rm b}=0.1520956(9)1$\thinspace d ($S/N=5.0$, $P_{\rm b}/P_1=0.634$) that do not fit the radial mode scenario. $P_{\rm b}/P_1$ however well fits the new group of radial--non-radial double mode pulsators discussed in this paper, provided the star in question is in fact RR~Lyrae. No comb., no cont. Non-radial modes or blend.

\noindent {\bf OGLE-BLG-RRLYR-03647}. $P_1=0.2534656(7)$\thinspace d (non-stationary). In addition we find $P_{\rm a}=0.262977(1)$\thinspace d ($S/N=8.8$). Period too close to 1O to be radial. No comb., no cont. Non-radial mode, Blazhko effect or blend.

\noindent {\bf OGLE-BLG-RRLYR-05550}. $P_1=0.24910497(3)$\thinspace d (non-stationary), $P_{\rm X}=0.1533192(5)$\thinspace d. In addition we find $P_{\rm a}=0.284199(2)$\thinspace d ($S/N=4.1$). $P_1/P_{\rm a}=0.877$ does not fit the radial mode scenario. No comb., no cont. Non-radial mode or blend. 

\noindent {\bf OGLE-BLG-RRLYR-06085}. $P_1=0.3068836(2)$\thinspace d. $P_{\rm X}=0.1881870(5)$\thinspace d. In addition we find $P_{\rm a}=0.1602604$\thinspace d ($S/N=4.5$), so $P_{\rm a}/P_1=0.522$ does not fit the radial mod scenario. No comb., no cont. Non-radial modes or blend.

\noindent {\bf OGLE-BLG-RRLYR-06345}. $P_1=0.20143783(4)$\thinspace d (non-stationary). In addition we find $P_{\rm a}=0.1897067(2)$\thinspace d ($S/N=13.2$) and next its very close, but well resolved companion $P_{\rm b}=0.1898976(4)$\thinspace d ($S/N=8.0$). These frequencies are too close to 1O and to each other to correspond to radial modes. No comb., no cont. Non-radial modes or blend.

\noindent {\bf OGLE-BLG-RRLYR-07486}. $P_1=0.31702951(4)$\thinspace d (non-stationary). In addition we find a longer period $P_{\rm a}=0.398583(3)$\thinspace d ($S/N=5.1$, non-stationary). Period ratio ($P_1/P_{\rm a}=0.795$) too high for the F+1O pulsator. It would fit the 1O+2O scenario provided that $P_1$ in fact corresponds to 2O. This possibility ruled out based on the shape of the light curve -- of large amplitude, described with sixth order Fourier series and typical for RRc stars.  No comb., no cont. Non-radial mode or blend.

\noindent {\bf OGLE-BLG-RRLYR-08125}. $P_1=0.27681180(6)$\thinspace d (non-stationary), $P_{\rm X}=0.169510(2)$\thinspace d. In addition we find $P_{\rm a}=0.149609(2)$\thinspace d ($S/N=6.3$) and then weak signal at $P_{\rm b}=0.2166949(7)$\thinspace d ($S/N=4.4$). Period ratios $P_{\rm a}/P_1=0.540$ and $P_{\rm b}/P_1=0.783$ do not fit the radial mode scenario.  No comb., no cont. Non-radial modes or blend.

\noindent {\bf OGLE-BLG-RRLYR-08463}. $P_1=0.34095554(3)$\thinspace d. In addition we find $P_{\rm a}=0.2137228(1)$\thinspace d ($S/N=12.6$) and its second and possibly third (weak) harmonic. Period ratio $P_{\rm a}/P_1=0.627$ fits the range in which we search for additional non-radial mode in this study. However large amplitude of additional signal and presence of its harmonics are not-typical for non-radial mode. Despite large amplitude, there are no combination frequencies. It is most likely a blend.

\noindent {\bf OGLE-BLG-RRLYR-09134}. $P_1=0.29693217(4)$\thinspace d, $P_{\rm X}=0.1824339(7)$\thinspace d. In addition we find a signal at longer period $P_{\rm a}=0.450203(3)$\thinspace d ($S/N=7.4$). $P_1/P_{\rm a}=0.660$ does not fit the F+1O scenario and light curve clearly indicate that $P_1$ must correspond to 1O. No comb., no cont. Non-radial mode or blend.

\noindent {\bf OGLE-BLG-RRLYR-09366}. $P_1=0.262936(2)$\thinspace d. In addition we find $P_{\rm a}=0.1372251(3)$\thinspace d ($S/N=9.1$) and next $P_{\rm b}=0.274405(3)$\thinspace d ($S/N=5.6$). $P_{\rm a}/P_1=0.522$ and $P_1/P_{\rm b}=0.958$. These period ratios do not fit the radial mode scenario. No comb., no cont. Non-radial modes or blend.

\noindent {\bf OGLE-BLG-RRLYR-10936}. $P_1=0.2056296(1)$\thinspace d. In addition we find $P_{\rm a}=0.1145288(3)$\thinspace d ($S/N=8.9$) and next $P_{\rm b}=0.0933232(3)$\thinspace d ($S/N=7.4$). $P_{\rm a}/P_1=0.557$ and $P_{\rm b}/P_1=0.454$. These period ratios do not fit the radial mode scenario. No comb., no cont. Non-radial modes or blend.

\noindent {\bf OGLE-BLG-RRLYR-11726}. $P_1=0.3019463(5)$\thinspace d, $P_{\rm X}=0.185382(3)$\thinspace d. In addition we find $P_{\rm a}=0.176021(1)$\thinspace d ($S/N=4.4$, $P_{\rm a}/P_1=0.583$). No comb., no cont. Non-radial mode or blend.

\noindent {\bf OGLE-BLG-RRLYR-12345}. $P_1=0.3280316(1)$\thinspace d. Its identification as RRc star is uncertain. No harmonic of $P_1$ is detected. We find at least three additional, close, but well resolved periodicities, all with $S/N$ well above 5. These include $P_{\rm a}=0.3447924(4)$\thinspace d, $P_{\rm b}=0.3421682(6)$\thinspace d, $P_{\rm c}=0.315773(1)$\thinspace d. No comb., no cont. A blend or multi-mode non-radial pulsator.

\bsp

\label{lastpage}


\begin{thebibliography}{99}
\bibitem[\protect\citeauthoryear{Benk\H{o} et al.}{2010}]{benko10} Benk\H{o} J.M., Kolenberg K., Szab\'o R. et al., 2010, MNRAS, 409, 1585
\bibitem[\protect\citeauthoryear{Benk\H{o} et al.}{2014}]{benko14} Benk\H{o} J.M., Plachy E., Szab\'o R. et al., 2014, ApJ Suppl. Ser., 213, 131
\bibitem[\protect\citeauthoryear{Chadid}{2012}]{chadid} Chadid M., 2012, A\&A, 540, A68
\bibitem[\protect\citeauthoryear{Dziembowski}{2012}]{wd12} Dziembowski W., 2012, Acta Astron., 62, 323
\bibitem[\protect\citeauthoryear{Dziembowski \& Smolec}{2009}]{wdrs09} Dziembowski W., Smolec R., 2012, AIP Conf. Proceedings, 1170, 83
\bibitem[\protect\citeauthoryear{Gruberbauer et al.}{2007}]{aqleo} Gruberbauer M., Kolenberg K., Rowe J. et al., 2007, MNRAS, 379, 1498
\bibitem[\protect\citeauthoryear{Jurcsik et al.}{2009}]{jurcsik} Jurcsik J., S\'odor A., Szeidl B. et al., 2009, MNRAS, 400, 1006
\bibitem[\protect\citeauthoryear{Mizerski}{2003}]{mizerski} Mizerski T., 2003, Acta Astron., 53, 307
\bibitem[\protect\citeauthoryear{Moskalik}{2013}]{pam13} Moskalik P., 2013, in: J.C. Su\'arez, R. Garrido, L.A. Balona, \& J. Christensen-Dalsgaard (eds.), Stellar Pulsations: Impact of New Instrumentation and New Insights, Astrophysics and Space Sci. Proc., Vol. 31 (Berlin, Heidelberg: Springer-Verlag), p. 103
\bibitem[\protect\citeauthoryear{Moskalik}{2014}]{pam14} Moskalik P., 2014, IAUS, 301, 249
\bibitem[\protect\citeauthoryear{Moskalik \& Ko\l{}aczkowski}{2009}]{mk09} Moskalik P., Ko\l{}aczkowski Z., 2009, MNRAS, 394, 1649
\bibitem[\protect\citeauthoryear{Moskalik et al.}{2013}]{pam_rrc} Moskalik P., Smolec R., Kolenberg K. et al., 2013, in: J.C. Su\'arez, R. Garrido, L.A. Balona, \& J. Christensen-Dalsgaard (eds.), Stellar Pulsations: Impact of New Instrumentation and New Insights, Astrophysics and Space Sci. Proc., Vol. 31 (Berlin, Heidelberg: Springer-Verlag), poster No 34 (arXiv:1208.4251)
\bibitem[\protect\citeauthoryear{Moskalik et al.}{2014}]{pamsm14} Moskalik P., Smolec R., Kolenberg K. et al., 2014, MNRAS, submitted
\bibitem[\protect\citeauthoryear{Nagy \& Kov\'acs}{2006}]{nagy} Nagy A., Kov\'acs G., 2006, A\&A, 454, 257
\bibitem[\protect\citeauthoryear{Olech \& Moskalik}{2009}]{om09} Olech A., Moskalik P., 2009, A\&A, 494, L17
\bibitem[\protect\citeauthoryear{Pietrukowicz et al.}{2013}]{pietruk} Pietrukowicz P., Dziembowski W., Mr\'oz P. et al., 2013, Acta Astron., 63, 379
\bibitem[\protect\citeauthoryear{Poleski et al.}{2010}]{poleski} Poleski R., Soszy\'nski I., Udalski A. et al., 2010, Acta Astron., 60, 1
\bibitem[\protect\citeauthoryear{Skarka}{2013}]{skarka} Skarka M., 2013, A\&A, 549, 101
\bibitem[\protect\citeauthoryear{Smith}{1995}]{smith} Smith H.A., 1994, RR~Lyrae stars, Cambridge Astrophysics Series, Vol. 27, Cambridge University Press
\bibitem[\protect\citeauthoryear{Smolec}{2014}]{smolec14} Smolec R., 2014, IAUS, 301, 265
\bibitem[\protect\citeauthoryear{Smolec \& Moskalik}{2008}]{sm08} Smolec R., Moskalik P., 2008, Acta Astron., 58, 193
\bibitem[\protect\citeauthoryear{Smolec \& Moskalik}{2010}]{sm10} Smolec R., Moskalik P., 2010, A\&A, 524, A40
\bibitem[\protect\citeauthoryear{Smolec et al.}{2012}]{blher} Smolec R., Soszy\'nski I., Moskalik P. et al., 2012, MNRAS, 419, 2407
\bibitem[\protect\citeauthoryear{Soszy\'nski et al.}{2008}]{ogle_freaks} Soszy\'nski I., Poleski R., Udalski A., et al., 2008, Acta Astron., 58, 153
\bibitem[\protect\citeauthoryear{Soszy\'nski et al.}{2009}]{ogle_rr_lmc} Soszy\'nski I., Udalski A., Szyma\'nski M.K. et al., 2009, Acta Astron., 59, 1
\bibitem[\protect\citeauthoryear{Soszy\'nski et al.}{2010}]{ogle_rr_smc} Soszy\'nski I., Udalski A., Szyma\'nski M.K. et al., 2010, Acta Astron., 60, 165
\bibitem[\protect\citeauthoryear{Soszy\'nski et al.}{2011}]{ogle_rr_blg} Soszy\'nski I., Dziembowski W., Udalski A., et al., 2011, Acta Astron., 61, 1
\bibitem[\protect\citeauthoryear{S\"uvegas et al.}{2012}]{sdss} S\"uvegas M., Sesar B., V\'aradi M. et al., 2012, MNRAS, 424, 2528
\bibitem[\protect\citeauthoryear{Szab\'o}{2014}]{szabo14} Szab\'o R., 2014, IAUS, 301, 241
\bibitem[\protect\citeauthoryear{Szab\'o et al.}{2014}]{szabo_corot} Szab\'o R., Benk\H{o} J.M., Papar\'o M., 2014, A\&A in the press (arXiv:1408.0653)
\bibitem[\protect\citeauthoryear{Udalski et al.}{2008}]{ogleIII} Udalski A., Szyma\'nski M.K., Soszy\'nski I., Poleski R., 2008, Acta Astron., 58, 69




\end{thebibliography}
\end{document}